\documentclass[12pt]{iopart}
\usepackage{amssymb,graphicx}

\newcommand{\non}{\nonumber \\}

     \newcommand{\del}{\delta}
\newcommand{\eps}{\epsilon}

\newcommand{\kap}{\kappa}     \newcommand{\lam}{\lambda}

\newcommand{\vphi}{\varphi}   \newcommand{\ome}{\omega}

    \newcommand{\cF}{{\cal F}}
    \newcommand{\cH}{{\cal H}}

\newcommand{\cM}{{\cal M}}    
\newcommand{\cO}{{\cal O}}

\newcommand{\pa}{\partial}

\newcommand{\rar}{\rightarrow}

\newcommand{\hlf}{\frac{1}{2}}
\newcommand{\ove}[1]{\frac{1}{#1}}

\newcommand{\gsim}{ \lower .75ex \hbox{$\sim$} \llap{\raise .27ex \hbox{$>$}} }

\begin{document}
\begin{flushright}
\parbox[t]{2in}{CERN-PH-TH/2004-143 \\
CU-TP-1123 \\
MAD-TH-04-07\\
hep-th/0411217}
\end{flushright}

\title[Backreaction barely constrains short distance
effects in the CMB.]{Decoupling in an expanding universe:\\
backreaction barely constrains short distance
effects in the CMB.}

\author{Brian R. Greene ${}^{1,2}$,
Koenraad Schalm ${}^1$,
Gary Shiu ${}^3$,
and Jan Pieter van der Schaar ${}^4$}

\address{
${}^1$ Institute for Strings, Cosmology and Astroparticle Physics  \& \\
     Department of Physics, Columbia University, New York, NY 10027}
\address{
${}^2$
     Department of Mathematics, Columbia University, New York, NY
     10027}
\address{
${}^3$
     Department of Physics,
     University of Wisconsin,
     Madison, WI 53706}
\address{
${}^4$
     Department of Physics, CERN, Theory Division, 1211 Geneva
     23}

\ead{ \mailto{greene@phys.columbia.edu}, \mailto{kschalm@phys.columbia.edu},\\
\mailto{shiu@physics.wisc.edu}, \mailto{jan.pieter.van.der.schaar@cern.ch.}}

\begin{abstract}
We clarify the status of transplanckian effects on the cosmic microwave
background (CMB) anisotro\-py. We do so using
the boundary effective action formalism of {\tt
  hep-th/0401164} which accounts quantitatively
for the cosmological vacuum ambiguity. In this formalism we can
clearly 1) delineate the validity of cosmological effective
actions in an expanding universe. The corollary of the initial
state ambiguity is the existence of an earliest time. The
inability of an effective action to describe physics before this
time demands that one sets initial conditions on the earliest time
hypersurface. A calculation then shows that CMB anisotropy
measurements are generically sensitive to high energy corrections to
the initial conditions. 2) We compute the one-loop contribution
to the stress-tensor due to high-energy physics corrections to an
arbitrary cosmological initial state. We find that
phenomenological bounds on the backreaction do not lead to strong
constraints on the coefficient of the leading boundary irrelevant
operator. Rather, we find that the power spectrum itself is the
quantity most sensitive to initial state corrections. 3) The computation of the one-loop
backreaction confirms arguments that irrelevant corrections to the
Bunch-Davies initial state yield non-adiabatic vacua characterized
by an energy excess at some earlier time. However, this excess
only dominates over the classical background at times before the
`earliest time' at which the effective action is valid. We
conclude that the cosmological effective action with boundaries is
a fully self-consistent and quantitative approach to
transplanckian corrections to the CMB.
\end{abstract}

\submitto{JCAP}

\maketitle
\setcounter{footnote}{0}
\section{Introduction}
\label{sec:introduction}

During the past few years, the prospect of probing ultra high energy
physics (on the order of the string/Planck scale) through precision
cosmological observations has received much attention. While more
traditional accelerator based approaches have the capacity to probe
string physics only if the string scale proves to be in the TeV
range, the enormous stretching of distance scales in inflationary
scenarios may well turn the cosmos into a {\it Planck-scale}
microscope: In all but the most conservative of inflationary models,
early-universe Planck scale physics is stretched by cosmological
expansion to scales relevant to the production of CMB temperature
fluctuations \cite{window, window2}. This suggests the tantalizing
possibility that measurements of CMB anisotropies may be a window
onto Planck scale physics.

The essential quantitative question facing this program is: How
large are the contributions coming from string/Planck scale physics,
and are they large enough to be observed? Many authors, using a
variety of approaches, have reached a range of conclusions. The most
optimistic estimates have calculated transplanckian contributions to
the CMB power spectrum ranging from $O(1)$
to $O(H/M_{\rm string})$ \cite{egks, Shiu:2002kg, Chu:2000ww, danielsson, Burgess:2003zw}
with $H$ the Hubble scale at the end of inflation, while more
pessimistic
estimates \cite{cos-vac-SK} suggest that the
maximum contributions are order $O((H/M_{\rm string})^2)$ (see further
\cite{cos-vac}). With a
favorable value of $H/M_{\rm string} \sim 10^{-2}$, the former
contribution would be on the edge of visibility \cite{Easther:2004vq},
while the latter
contributions (without anomalously large coefficients) would fall
below the limit of cosmic variance. Achieving consensus on the
likely size of transplanckian corrections to the CMB power spectrum
is thus an issue of great importance.

In this paper, we focus on transplanckian effects on the CMB
coming from the well known and much studied ambiguity in choosing
a vacuum state in a nontrivial cosmological space-time. By way of
brief history, in \cite{egks} a combination of such a modified
vacuum state together with string-motivated modifications to the
dynamical evolution equations were studied, with the conclusion
that order $H/M_{string}$ corrections are generic. In
\cite{danielsson} and the third article of \cite{egks}, it was
emphasized that a modified vacuum state together with {\it
standard} evolution equations would also give rise to generic
$H/M_{string}$ corrections, thus clarifying and simplifying the
origin of transplanckian contributions of this magnitude. However,
the papers \cite{cos-vac-SK}, employing effective field theory
techniques, argued that the standard vacuum state (Bunch-Davies
boundary conditions) would yield the far smaller $(H/M_{\rm
string})^2$ contributions, and that, moreover, any other choice
for the boundary conditions would fall prey to uncontrollable
backreaction in the absence of fine tuning. Inspired in part by
\cite{Burgess:2003zw}, in \cite{Schalm:2004qk} a quantitative
formalism for studying the ambiguity in cosmological initial
conditions in the context of boundary effective field theory was
put forward. It was argued that the integrating out of high energy
modes {\it generically} modifies the boundary conditions of the
resulting low energy effective field theory. Importantly, the size
of the corrections was found to be of order $H/M_{string}$,
agreeing with the early papers on high energy effects on the
vacuum choice, but with calculations carried out in a controlled
approximation. Moreover, the boundary effective action provided
new counterterms that would render the gravitational backreaction
under control for choices other than Bunch-Davies boundary
conditions. Subsequently, an order of magnitude estimate of the
now manifestly finite backreaction \cite{Porrati:2004gz} argued
that the observed slow-roll period of inflation significantly
bounds the size of potential transplanckian signatures in the CMB
anisotropy.

In this paper, armed with the quantitative boundary effective
field theory framework, we extend the work of \cite{Schalm:2004qk}
and \cite{Porrati:2004gz} by carefully calculating the
gravitational backreaction. Seeking fully unambigous
backreaction constraints --- backreaction contributions that are not
subject to any renormalization ambiguities and hence are fully
interpretable in the effective field theory framework --- we find
robust conclusions on the size of transplanckian CMB corrections.
In particular, the existence of new
counterterms has the potential to subdue the order of magnitude estimate of the
one-loop backreaction to the point that there are no bounds of
consequence on potential corrections to the observed power
spectrum {for $H/M
> 10^{-4}$}. Without additional information from a proposed UV-completion of our effective field theory
description --- one that would allow a more refined estimate of the
gravitational backreaction --- we conclude that backreaction
constraints are under control for parameter ranges that can, in
principle, yield observable consequences for the cosmic microwave
background radiation. Barring a cosmological observation
more sensitive to high-energy corrections to the initial state,
the theoretical window of opportunity to observe short distance
physics in the CMB is thus open.

\subsection{Summary of Results}
\label{sec:summ}

In section \ref{sec:beft} we review
the results of \cite{Schalm:2004qk} that
provide a new formalism to address the initial state ambiguity in
cosmological space-times in a coherent
Lagrangian effective field theory description. The initial state can
be encoded in a (space-like) boundary action at an a priori arbitrary initial
time $t_0$.
From the standard
Hamiltonian perspective, this corresponds to scenarios
where the Bogoliubov coefficients parameterizing the initial
state ambiguity are allowed to vary with co-moving momentum $k$.
The distinct
advantage of the Lagrangian
effective field theory approach
is
that it is the natural framework to consider effects of high
 energy physics (see e.g. \cite{Polchinski:1983gv}). It
 allows one to neglect momentum modes beyond the high
 energy cut-off in a manifestly consistent way and it provides a {\em quantitative} understanding of
 effects due to UV physics. Specifically, it recasts
the momentum-dependence of the initial state into a well-defined
expansion in irrelevant operators parameterizing unknown
high-energy physics.

The leading effect of high-energy physics arises from these
irrelevant corrections to the boundary action rather than from
higher derivative corrections in the bulk (which were analyzed in
\cite{Shiu:2002kg,cos-vac-SK}). The leading irrelevant operator
on the three-dimensional boundary is dimension four. Corrections
to the inflationary power spectrum therefore behave parametrically
as $H/M$, which can conceivably be as high as 1\%. This renders
them potentially observable in future CMB measurements.

\subsubsection*{An earliest time}

As we will discuss in \ref{sec:an-earliest-time}, an immediate
consequence of the effective field theory framework in a
cosmological context --- of demanding that all the momentum modes
under consideration are {\it always} smaller than the physical
cutoff scale --- is the existence of an `earliest time' before which
the effective field theory breaks down. This earliest time
depends on the momentum scale of interest and
roughly corresponds to $\ln(M/H)$ e-folds before horizon exit
of the smallest observable length scale in the CMB. It is the
natural location to set the initial conditions.\footnote{In practice
  this natural proposition raises a question.
  For optimistic inflationary scenarios with $M/H \sim
  10^{2}$, the lowest modes observable in the CMB --- four orders of
  magnitude below the smallest observable mode --- have already exited
  the horizon $\ln(10^2)$ e-folds earlier. Usually horizon exit is
  interpreted as modes ceasing to be dynamical and being frozen
  out. If this interpretation is truly correct,
  how can these modes account for low multipole CMB fluctuations?
  At face value, 1) the imposition of boundary conditions at a fixed
`earliest' time, 2) the
non-uniform times $a)$ when each individual mode is of the order of the
cut-off and $b)$
when it exits the horizon, and 3) the wish to describe the
full range of observed
modes in the CMB at once, are in tension with each other.
  For the power spectrum we can circumvent this conflict as a linear
analysis is
  sufficient. Modes do not interact, and one can set the initial
  conditions for each mode at different times. For
  an interacting field theory it is open question ,
  however, how to resolve the tension between 1), 2) and 3).}

By doing so, we find that the current measurements of the power
spectrum of CMB fluctuations already provide a strong bound on
the {\em coefficient} $\beta$ of the leading irrelevant operator
in the boundary effective action if $H/M > 10^{-4}$.
This is directly related to the
fact that the initial state deduced from the boundary effective
field theory will generically break de Sitter scale invariance,
(i.e. the Bogoliubov coefficients depend on the co-moving momentum
scale in Hamiltonian language).
Roughly put,
the effect of the leading irrelevant boundary operator adds a
linear component $\del P = \beta k/a_0 M$ to the power spectrum,
where
 $a_0$ is the scale factor at the `earliest time'.  The
 change to the power spectrum is thus parametrically controlled by
 $k/aM \sim H/M$,
 but at the high end of the spectrum it can be significantly
 larger than that (figure \ref{fig:2} in section
 \ref{sec:an-earliest-time}).
This linear growth is not present in the observed CMB spectrum; it
  is nearly scale invariant over the full
observed range.
This rules out a coefficient $\beta$ much larger than $0.1$.

We see that an earliest time has as primary consequence that
inflation's usual classical independence of initial conditions is
modified as the quantum mechanical boundary conditions affect
late-time physics.
In particular, if the modes
responsible for the CMB perturbations that we can measure today
exited the horizon a sufficiently small number of e-foldings after
the earliest time hypersurface, then sensitivity to initial
conditions can persist.
Nevertheless, the cosmological dynamics that result from the
classical background motion of the scalar field
are still expected to be independent of the initial conditions
\cite{bozza}.

\subsubsection*{Backreaction constraints}

Naturally, all other --- measured --- cosmological quantities will
also be affected by the irrelevant boundary operators and
observability therefore hinges on whether other phenomenological
constraints are mild enough to allow a 1\% change to the power
spectrum.
 In particular, \cite{Porrati:2004gz} argued that an order of magnitude estimate of
the gravitational backreaction yields constraints on
transplanckian calculations that are quite
significant.\footnote{That backreaction effects in this context
could be important was also emphasized in \cite{Tanaka:2000jw} (see
also \cite{Lemoine:2001ar} and the recent articles \cite{Giovannini:2003it,Brandenberger:2004kx})). Other phenomenological constraints on
initial state modifications have been discussed in
\cite{Starobinsky:2002rp}.
More formal arguments against the use
of non-standard initial states can be found in \cite{cos-vac-SK,
  cos-vac-crit}.}
We compute here the gravitational backreaction in detail. The
resulting perturbative bound on the coefficient $\beta$ of the
leading irrelevant boundary operator,
\begin{eqnarray}
\label{eq:260a} |\beta|^2 &\ll& (12 \pi)^2 \, \left({M_p^2 H_0^2
\over M_{string}^4}\right)~,
\end{eqnarray}
plus the constraints from the
observed inflationary slow-roll parameters $\eps_{observ},
\eta_{observ}$
\begin{eqnarray}
\label{eq:11}
|\beta|^2 &\lesssim& 2 \, (6 \pi)^2 \, |\epsilon_{observ}| \, \left( {M_p^2 H_0^2 \over M_{string}^4} \right)~, \\
\label{eq:12}
|\beta|^2 &\lesssim& (6 \pi)^2 \, |\epsilon_{observ}| \, |\eta_{observ}| \, \left( {M_p^2 H_0^2 \over M_{string}^4} \right)~,
\end{eqnarray}
entail relatively mild backreaction constraints. For typical but
optimistic values for $H \sim 10^{14}$ GeV, the scale of new
physics $M_{string} \sim 10^{16}$ GeV, and the reduced Planck mass
$M_p \sim 10^{19}$ GeV, they allow a significant observational
window of opportunity. The mildness results from the fact that 
the first unambiguous contribution to the backreaction is only at
{\em second} order in the irrelevant correction (This had earlier
been argued by Tanaka \cite{Tanaka:2000jw}. Indeed compared to the
order of magnitude estimate \cite{Porrati:2004gz} the above three
equations are effectively the same with $\beta^2$ substituted for
$\beta$.) The backreaction due to the first order correction,
though not zero, is essentially localized on the boundary and
therefore subject to the subtraction prescription utilized to
renormalize the theory. 
The localization is a consequence of the
highly oscillatory nature of the first order power-spectrum. When
integrated, all contributions cancel except in a
neighborhood of the boundary. In the context of effective
field theory --- the only context we consider in this paper --- such
terms are subject to renormalization via boundary counterterms and
hence their contribution is ambiguous. By contrast, the second
order effect which remains and dominates is the `time-averaged'
energy stored in the oscillatory behavior itself, which
gives a contribution whose scale is controlled by $H$ and hence is
clearly physical. In particular, this contribution grows as the
square of the amplitude rather than linearly, and it is this which
accounts for the appearance of $|\beta|^2$ rather than $|\beta|$
in eqs. (\ref{eq:260a})-~(\ref{eq:12}) above.

The bounds on the coefficient $\beta$ due to the one-loop
backreaction are in fact so mild that they are superseded by the
direct sensitivity of the power spectrum.
The aforementioned
existence of an earliest time and its concomitant bound on $\beta
\leq 0.1$ implies that backreaction poses {\em no} constraints
if $H/M$ is large enough (figure \ref{fig:2} in section
\ref{sec:an-earliest-time}).
The bounds on $\beta$ from backreaction are weaker than the
direct `search' upper bound from the power-spectrum, allowing for
the possibility of non-trivial implications for CMB observations.

\subsubsection*{Energy and adiabaticity of the initial state}

The calculation of irrelevant contributions to the one-loop
backreaction does make clear that the presence of irrelevant
corrections to the Bunch-Davies boundary action must amount to an
extra vacuum energy presence in the space-time. The non-localized
second order contribution to the backreaction redshifts away as
any other energy density. Qualitatively this indicates that the
effective action has a limited range of applicability to the far
past to where the quantum vacuum energy overwhelms the classical
one, as has been argued in \cite{cos-vac-SK}. The boundary action
formalism allows a quantitative calculation of this range of
applicability. One finds that the excess energy stored in the
initial state only dominates {\em before} the earliest time where
the effective action ceases to be valid. Within this formalism we
are therefore intrinsically unable to answer what happens before
that moment. This is not to say that the criticism that the excess
energy could ruin inflation a few $e$-folds further back is
without merit. This is certainly an issue, and one that would need
to be addressed by fully fundamental description. But within the
framework of a low energy effective action it is an issue that we
need not, and in fact, cannot address: by definition one cannot
answer any questions outside the effective field theory's range of
validity. The whole framework is therefore self-consistent, and it
crucially hinges on the existence of an earliest time. Within this
range of validity backreaction never gets out of hand.

Beyond consistency, one might worry about fine-tuning: How finely
do we have to tune the initial state to obtain observably large
transplanckian effects in the CMB without falling afoul of a
theoretical or phenomenological constraint? We will address this
briefly in the conclusion, section \ref{sec:conclusion}. A
fine-tuning requirement itself is no phenomenological constraint,
however, but a theoretical prejudice. It is not related to
consistency of the boundary effective action framework or
phenomenological bounds on potential short distance effects in the
CMB. For this reason we postpone a full accounting of fine-tuning
issues to a separate article.

Since energy is stored in the initial state, this does show that a
non standard choice of initial conditions necessarily corresponds
to a state differing from the adiabatic ground state of the
system. The irrelevant corrections to the initial state therefore
parameterize deviations from adiabaticity.

\medskip

The above demonstrates that the use of a boundary effective action
to parameterize the cosmological vacuum ambiguity, allows us to
{\em quantitatively and reliably} describe and compute high energy
corrections  to low energy and late time cosmological physics,
including those to the initial conditions. The calculations
confirm qualitative estimates about phenomenological constraints
on irrelevant non-adiabatic corrections to the initial state, with
important caveats. The backreaction is only significantly affected
at second order. Hence 1) the size of phenomenologically allowed
irrelevant corrections is significantly larger than dimensional
analysis would suggest and 2) for $H/M>10^{-4}$ the
inflationary power spectrum of
fluctuations itself is the most sensitive measurement of
irrelevant corrections to the initial conditions. In particular
the corrections to the power spectrum  are allowed to be of order
1\%  without significantly disturbing the inflationary background.
As a consequence backreaction poses little constraint and initial
state corrections due to the irrelevant operators could be large
enough to be observed. Indeed given that the backreaction bounds
are so weak, the observed scale invariance of the CMB bounds
$\beta \lesssim 0.1$ {\em directly.}

\section{Initial states in effective field theory and initial states:
a brief
review}
\label{sec:beft}

For simplicity we will consider an external scalar field in
conformal gauge de Sitter.
This situation is
applicable for tensor-fluctuations in the CMB. We expect our
results to carry over to the dominant (and measured)
scalar-fluctuations without significant change beyond the known
amplitude magnification.

In the formalism of \cite{Schalm:2004qk} the cosmological vacuum
choice ambiguity is captured by a boundary action at an arbitrary
initial time $t_0$. A specific choice of boundary couplings
corresponds to a specific choice of initial state. For $\lambda
\phi^4$ theory on a semi-infinite space\footnote{We consider
$4$-dimensional Lorentzian space-times and introduce a space-like
$3$-dimensional boundary, allowing the boundary dynamics to have a
natural interpretation in terms of how they affect the initial
state. Working with effective Lagrangians we will implicitly
assume that our results can be obtained by a Wick rotation from
Euclidean space.}
\begin{eqnarray}
  \label{eq:10}
  S_{bulk} = \int_{t_0 \leq t\leq \infty} \hspace{-0.4in}d^3 x dt\,
  \sqrt{g} \left( - \frac{1}{2} \partial_{\mu} \phi\pa^\mu\phi -
  \frac{m^2}{2}\phi^2 -\frac{\lam}{4!}\phi^4 \right)~,
\end{eqnarray}
with the following boundary interactions
\begin{eqnarray}
  \label{eq:20}
  S_{boundary} = \oint_{t=t_0} d^3x \sqrt{g_{\mu\nu}\pa_ix^{\mu}\pa_jx^{\nu}}\left( - \frac{\kappa_0}{2}\phi^2\right) ~,
\end{eqnarray}
variation of the bulk and boundary action\footnote{The variations on
  the boundary are arbitrary; otherwise we would be imposing Dirichlet boundary
conditions.} yields the
usual bulk equations of motion {\it plus} a boundary term that
  vanishes when the normal derivative of $\phi$ obeys
\begin{eqnarray}
  \label{eq:30}
  \left. \partial_n\phi \right|_{t=t_0} = -\kappa_0 \, \phi(t_0) ~.
\end{eqnarray}
This initial condition on the (classical) fields encodes the quantum state
or equivalently the scalar field wave-functional at $t=t_0$. The specific value of
$\kappa_0$ uniquely determines the boundary condition: for $\kappa_0=0$ we have Neumann
boundary conditions, for $|\kappa_0| \rightarrow \infty$ we find a Dirichlet
boundary condition and for finite $\kappa_0$ some mixture of the two.
For constant $\kappa$ the boundary action is renormalizable. One can
impose different boundary conditions for different (spatial) momentum modes by
choosing a momentum dependent effective $\kap_{eff}(k)$. Expanding
around $k=0$ following the precepts of effective actions, returns the
power-counting renormalizable (relevant and marginal) couplings $\kap_0$ and $\kap_1(k) = \alpha
|k|$ plus a set of
nonrenormalizable higher derivative (irrelevant) operators $\kap_n(k)
= \beta_{n-1} |k|^{n} /M^{n-1}$. These
represent our ignorance of (very) high energy physics beyond the
physical cut-off scale $M$.
In a flat Minkowski $\kap_{eff}(k)$ is uniquely determined by Lorentz
symmetry: $\kap_{Mink}(k)  = -i\ome(k)$. In cosmological settings the
absence of Lorentz symmetry allows {\em a priori} more general initial
conditions, {including irrelevant corrections due to unknown
  high-energy physics.}\footnote{Of course, the absence of Lorentz
  symmetry is by itself not enough to create a vacuum/initial state
  ambiguity. If there exists a global timelike Killing vector the
  vacuum is uniquely determined \cite{Birrell:ix,Fulling}.}
In \cite{Schalm:2004qk} all the leading dimension four
irrelevant operators respecting the homogeneity and isotropy of FRW
cosmologies were
constructed and analyzed. In this letter we will focus our attention
on one of them:\footnote{The one-loop backreaction, which is the goal
  of this letter,
is dominated
by the high-$k$ modes and all leading dimension $4$ operators reduce
to eq. (\ref{eq:40}) in the high $k$ limit. In the language of
  \cite{Schalm:2004qk} the coefficient $\beta$ equals $\beta =
  \beta_{\parallel}-\beta_c -\beta_{\perp}$.}
\begin{eqnarray}
  \label{eq:40}
  S_{bound}^{irr.op.} = \oint_{\eta=\eta_0} d^3x \, a_0 \, \left(
  -\frac{\beta}{2M}\partial^i\phi\partial_i\phi \right) ~.
\end{eqnarray}
Here $a_0\equiv a(\eta_0)=-1/H\eta_0$
is the de Sitter scale factor evaluated on
the boundary surface at conformal time $\eta=\eta_0$.
This specific irrelevant
operator is also the one analyzed in \cite{Porrati:2004gz}. It
leads to the following
${\cal O}(1/M)$ corrections to the relevant coupling $\kappa_0$
\begin{eqnarray}
  \label{eq:50}
  \kappa_{eff}(k) =\kappa_0 + \beta \, \frac{k^2}{a_0^2 \, M}~.
\end{eqnarray}
The dimensionless coefficient $\beta$ is in principle determined
by the UV completion of the theory and is expected to be of ${\cal
O}(1)$ if $M$ is the scale of new physics (although the
possibility of fine-tuned coefficients cannot be excluded, of
course). The phenomenological importance of a  $\beta \sim \cO(1)$
irrelevant correction is that it leads to corrections to the
inflationary power spectrum parametrically controlled by $H/M$,
that might be detectable in future CMB observations. Explicitly,
the change in the power spectrum due to the irrelevant operator
(\ref{eq:40}) is given by
\begin{eqnarray}
  \label{eq:3c}
  \frac{P+\del P}{P}(y_0) = \left(1 +
  \frac{\pi \beta H}{4M}\left(i\bar{\cH}_{3/2}^2(y_0)y_0^2+{\rm c.c}\right)\right)~,
\end{eqnarray}
where $y_0 \equiv k/a_0H$ is the physical momentum in units of the
horizon-size at the time where we set the initial conditions, and
$\cH_{3/2}(y_0)$ is the Hankel function.
(For details underlying this result we refer to
\cite{Schalm:2004qk}.)
Because these corrections
 will break the scale invariance of the inflationary power spectrum
 (see Figure \ref{fig:2}), a coefficient of $\cO(1)$ is in fact
 already directly excluded from the observed CMB
 spectrum.\footnote{Scale invariance as a characteristic of the
 inflationary power spectrum is especially emphasized in
 \cite{larsen}.}
Our aim here will be to show that the observed scale invariance
is truly a direct bound on
 the size of $\beta$ and not superseded by phenomenological bounds on $\beta$ due to gravitational
 backreaction.

\subsection{An earliest time in cosmological effective actions.\\ The
  inflationary power spectrum}
\label{sec:an-earliest-time}

Perturbative effective actions are intrinsically limited in their
range of validity to scales below the physical cut-off $M$. In
cosmological effective actions this means that the action is expected
to break down both for high scales and early times where all momenta
are blueshifted. A strength of the boundary effective action formalism
is that the momentum expansion is manifestly controlled by
two small parameters: the bulk action is controlled by $k/a(t)M$, the
boundary action by $k/a_0M$.
We immediately see that an FRW effective
action is only valid up to the 'scale'
\begin{eqnarray}
  \label{eq:2c}
 \mu_{phys}(t) \equiv \frac{\mu_{co}}{a(t)} = M~,
\end{eqnarray}
and only valid up to a `time'
\begin{eqnarray}
  \label{eq:1c}
  a_0 = \mu_{co}/M~.
\end{eqnarray}
where $\mu_{co}$ is the scale of
interest in units co-moving
momentum; i.e. the largest
comoving momentum mode we can see in today's CMB.
 We see here the mathematical manifestation of our
intuition that we can only trust low energy effective cosmological
theories up to the `Planck time'.

This `earliest time' is the logical place to locate the
boundary action to set the initial conditions. Doing so, we can
refine our analysis for which values of $\beta$ and $H/M$ changes in
the power spectrum are of the right order of magnitude to be
potentially observable.

Clearly the maximal change in the
power spectrum occurs for the largest possible value of $y_{0,max}=
k_{max,observed}/a_0H$. This is simply a consequence of the fact that
we are studying the effects of irrelevant operators whose size
increases with $\vec{k}$. Having realized the existence of an
earliest time $a_0 = k_{max}/M$, we set the initial conditions there
(we {\em cannot} choose an earlier
time with $a_0$ less than that; we could choose a larger value at a
later time). Hence $y_{0,max} = M/H$, i.e. by construction the highest
momentum mode observed in the power spectrum is `scaled' to
$y_{0,max}=M/H$. The observed CMB power spectrum stretches to four
orders of magnitude below that; it
therefore ranges from $10^{-4}y_0$ to $y_0$.

\begin{figure}[hp]
  \centerline{
\includegraphics[width=3.1in]{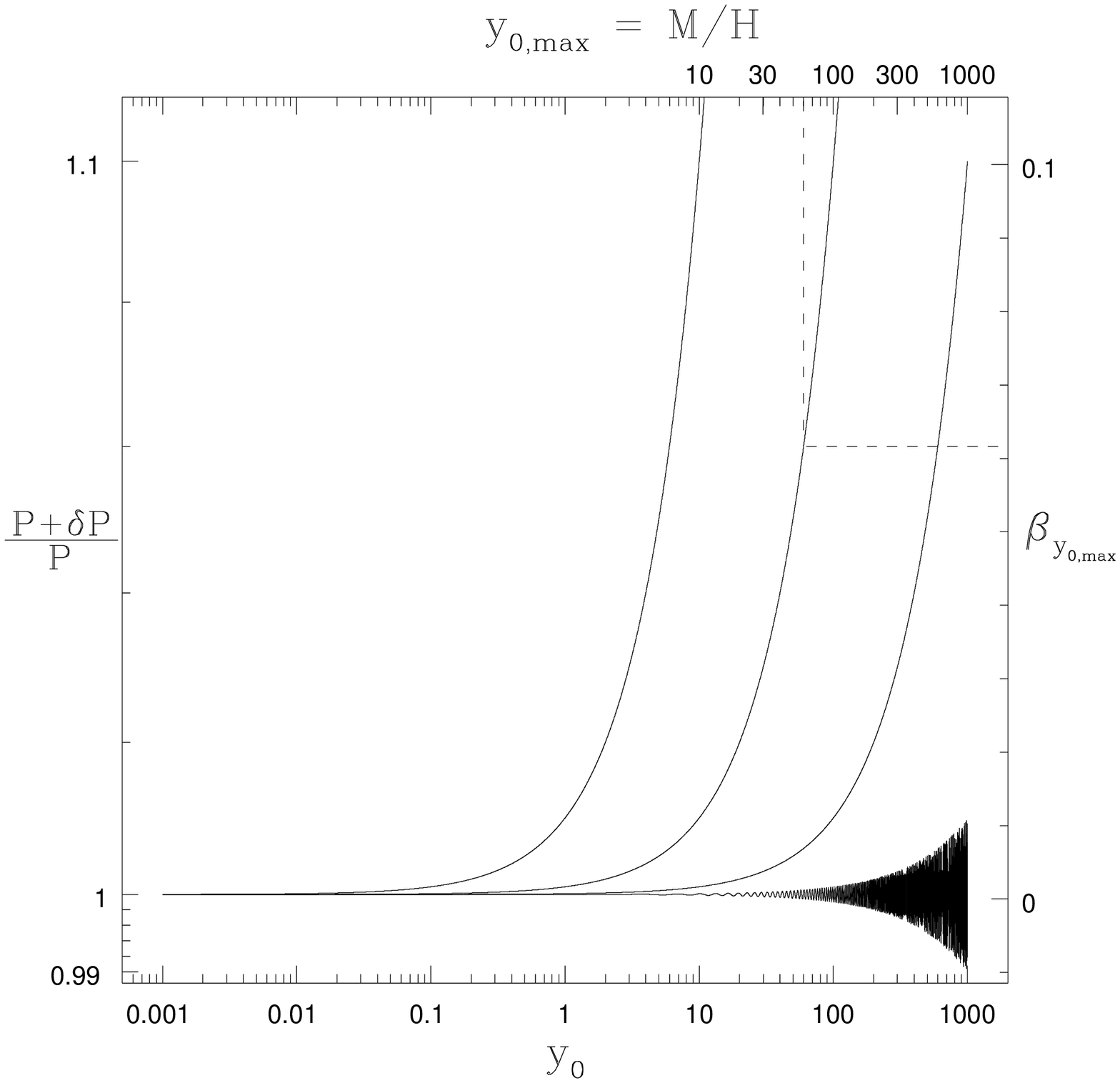}
\includegraphics[width=3.1in]{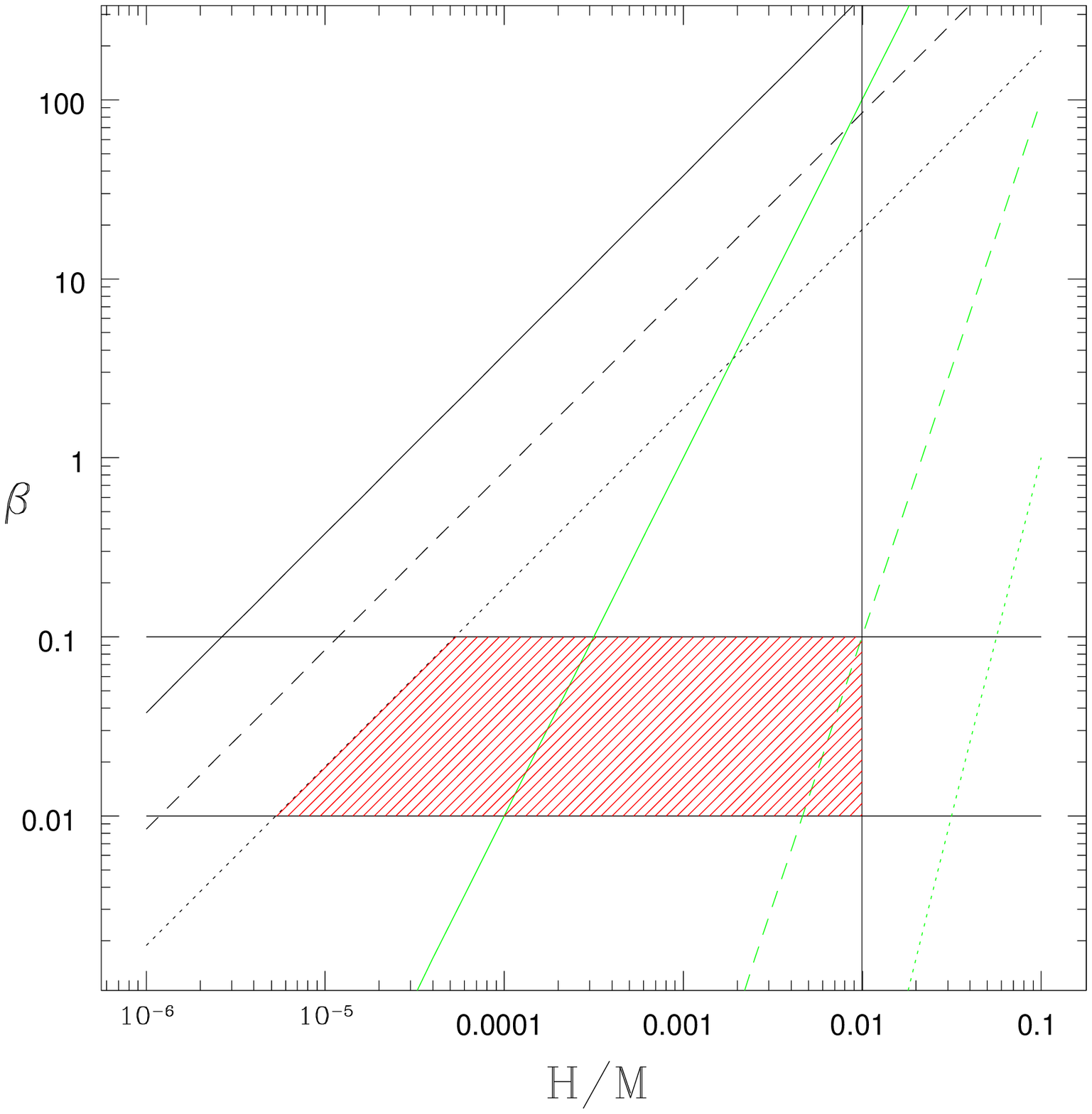}
}
  \caption{{\em A refined estimate of the sensitivity of the CMB to
      new physics.} \hfill {$\,$}
The left panel shows the change in the (amplitude of the)
power spectrum due to the
presence of the leading order irrelevant operator $\frac{\beta}{M}
(\pa_i\phi)^2$ as a function of
the physical momentum in units of the size of the horizon at the
earliest time. (Only for one specific choice is the full oscillatory
Bessel function behavior plotted.)
This graph should be read as follows. Given the scale
of new physics $M$ and the Hubble constant $H$ during inflation (or
more precisely at
the time when the highest mode $k_{max}$ of interest exits the horizon) the
earliest time up to which we can trust the effective action is
$y_{0,max}\equiv k_{max}/a_{0,min}H =M/H$
(see section \ref{sec:an-earliest-time}). Anything to the right of
$y_{0,max}$ should be discarded as untrustworthy. {The observed CMB
stretches to four orders of magnitude smaller momenta
from $10^{-4}y_{0,max}$ to $y_{0,max}$.} Precisely at
$y_{0,max}$ the change in the power spectrum is linearly dependent on
the value of $\beta$. The values of $M/H$ and $\beta$ corresponding to the
various curves can thus be read off from the intersection of the
plumblines to the upper and right axis.
The right panel shows an exclusion plot for $\beta$ as a function of
$H/M$. The $45^o$ lines correspond to the backreaction bounds derived
in section \ref{sec:backreacion} (continuous for zeroth order in slow
roll, dashed for first order in slow roll, dotted for second order in
slow roll). {The $60^o$ lines correspond to the order of magnitude
estimate for the backreaction \cite{Porrati:2004gz}; they are
equivalent to an estimate based on dimensional analysis.}
The upper horizontal line is an order of magnitude
estimation of the current error to which we have a nearly
scale invariant spectrum \cite{WMAP}. The lower horizontal line is an
order of magnitude estimate of the cosmic variance limitations of
resolution. Finally the vertical line denotes a maximal value of
$H/M$ consistent with observation using a value for $H$ extracted from the
allowed scalar/tensor ratio and $M \equiv 10^{16}$ GeV. This leaves the
shaded region as the {\em window of opportunity} to observe
transplanckian physics in the CMB.
}
\label{fig:2}
\end{figure}

For
this maximal value of $y_0$ we see that the change in the
power spectrum equals
\begin{eqnarray}
  \label{eq:2}
\frac{P^{dS}_{BD}+\del P}{P^{dS}_{BD}} (y_{0,max}) &=& 1+
\frac{\pi}{4}\frac{\beta H}{M} \left[ i\frac{M^2}{H^2}
  \bar{\cH}_{3/2}^2(M/H) + {\rm c.c.}\right] \non
&\simeq& 1 + \beta \sin(2M/H)~.
\end{eqnarray}
Note: though the change in the power spectrum is parametrically
$H/M$ as argued before (see eq. (\ref{eq:3c})), its maximal change
is quite independent of their values --- if one chooses
$a_0=k_{max}/M \leftrightarrow y_{0,max}=M/H$. For this specific
value of $y_0$ the change in the power spectrum is linearly
dependent on the size of the irrelevant operator $\beta$. We have
{drawn the change in the power spectrum} in figure \ref{fig:2}.
The explicit breaking of de Sitter scale invariance by the leading
irrelevant operator results in a linear momentum dependence of the
amplitude of the oscillatory correction to power spectrum. This
enhancement at high momentum means that for small values of
$\beta$ and only moderately large values of $M/H$ the change in
the power spectrum is far larger than the projected $1\%$
uncertainty in future measurements. We have a solid case that for
a large enough value of $H/M$ future CMB measurements are
sensitive to high energy physics through irrelevant corrections to
the initial conditions. Moreover, figure \ref{fig:2} clearly shows
that the current sensitivity with which the power spectrum is
measured already constrains the allowed values for $\beta$ and
$H/M$ in nature. A coarse extrapolation from the WMAP results
\cite{WMAP} indicates that the observed power spectrum is scale
invariant with an accuracy of around $10\%$. A value of $\beta
\sim 0.2$ and $H/M \sim 0.01$ would already imply a $20\%$ change
at the upper end of the power spectrum, inconsistent with the
data. This establishes the point of principle that the power
spectrum can be sensitive to irrelevant corrections. 
Recent data sensitivity studies confirm in much more
detail that in 
specific scenarios the contribution of physics beyond $M$ can be
disentangled from the data given a high-enough value of $H/M$ \cite{Easther:2004vq}.

{Other measurements,
however, in particular the near absence of gravitational
backreaction could} constrain the size of the irrelevant operator
beyond the direct measurement in the power spectrum. Indeed an
order of magnitude estimate indicates that this will be so
\cite{Porrati:2004gz}. {This is illustrated} in the second panel
of figure \ref{fig:2}. The remainder of this article will show
that a precise calculation of the gravitational backreaction
reveals that the leading order of magnitude result is subject to a
renormalization prescription. The physical second order result
imposes weaker constraints on the size of irrelevant corrections.
There is therefore a distinct {\em window of opportunity} to
measure short distance physics in the CMB: the shaded region in
figure \ref{fig:2}.

\subsection{Corrections to the Green's function}

To compute the one-loop correction to the gravitational
stress-tensor, we will need the Green's function for $\phi$ with
initial conditions set by the effective boundary coupling
$\kap_{eff}$ in eq. (\ref{eq:50}). Rather than adhering to the
Lagrangian formalism, involving the parameter $\kappa$ to encode
the initial state, it is instructive to translate back to the
Hamiltonian formalism in which the Bogoliubov coefficients $b(k)$
parameterize the initial state. We do so to make contact with the
standard approach in cosmology: details of the Lagrangian approach
are discussed in \cite{Schalm:2004qk}.\footnote{An extensive treatment
  of the
Hamiltonian approach
adiabatic order 4 vacua is forthcoming \cite{mottola-new}. Here, as
mentioned, we will be discussing non-adiabatic initial states.}

Introducing two independent,
homogeneous
solutions $\varphi_{\pm}(k,\eta)$ of the bulk
equations, the general solution is
\begin{eqnarray}
\label{eq:60}
\varphi_b(k,\eta) = \varphi_+ (k,\eta) + b(k) \, \varphi_- (k,\eta)~,
\end{eqnarray}
with $\varphi_- =
(\varphi_+)^*$ and normalized
according to the Klein-Gordon inner product
(see eq. (\ref{eq:110})).

Expanding the fields onto a basis of the
independent solutions, and promoting them to operators,
\begin{eqnarray}
\label{eq:80} \phi_b(x,\eta)=\int {d^3k \over (2\pi)^{3}} \left[
\hat{a}_b(k) \bar{\varphi}_{b}(k,\eta) + \hat{a}^\dagger_b(-k)
{\varphi}_{b}(-k,\eta)\right]\, e^{i\vec{k} \cdot \vec{x}}~,
\end{eqnarray}
one defines the vacuum as the state which is annihilated by
$\hat{a}_b(k)\, |b\rangle =0$. With respect to this vacuum, we
then construct a Green's function from the time-ordered product of
fields.
\begin{eqnarray}
  \label{eq:13}
  G_b(k;\eta_1,\eta_2) &\equiv& \langle
  b|T(\phi_b(k,\eta_1)\phi_b(k,\eta_2)|b \rangle
\non
&=&
  \frac{1}{(1-b\bar{b})}\left(\bar{\vphi}_{b}(k,\eta_1)\vphi_{b}(k,\eta_2)\theta(\eta_1-\eta_2)
  + (\eta_1 \leftrightarrow \eta_2)\right)~.
\end{eqnarray}
Demanding that the Green's function obeys the boundary condition
$$
\left.a(\eta_0)^{-1}\pa_{\eta} G_b(k;\eta,\eta_2)\right|_{\eta=\eta_0}
=-\kap G_b(k;\eta_0,\eta_2)
$$
inherited from eq. (\ref{eq:30}), one immediately realizes that this
implies that the mode-function $\vphi_{b}$ obeys the same boundary
condition as the field.\footnote{The Green's function is a solution to
  a second order differential equation and therefore requires two
  boundary conditions to determine it uniquely. The other boundary
  condition in this case is $\pa_nG|_{\eta= far~future} =
  -\bar{\kap}G$.  Chapter 6 of \cite{Birrell:ix} explains why
  for the dominant part of the gravitational backreaction we can
  choose the in- and out-states the same.
  How to impose two independent boundary conditions in
  the future and the past is explained in
  \cite{Schalm:2004qk}.}
It is then straightforward to deduce the following relation
between the parameters $\kappa(k)$ and $b(k)$
\begin{eqnarray}
\label{eq:70}
b_{\kap}(k) = - { \kappa(k) \varphi_{+,0} + \partial_n \varphi_{+,0} \over \kappa(k) \varphi_{-,0} +
\partial_n \varphi_{-,0}}~,~~~~\varphi_{\pm,0}\equiv \varphi_{\pm}(\eta_0)~.
\end{eqnarray}
The vacuum state $|b_{\kap}\rangle$ which is annihilated by
$\hat{a}_{b_{\kap}}(k)$
corresponds to the initial conditions in effective field
theory set by the boundary coupling parameter $\kappa(k)$.
The complicated relation between $b(k)$ and $\kappa(k)$
illustrates why the Hamiltonian framework is not well-suited to discuss
renormalization and effective actions. The Hamiltonian theory has no
natural expansion in irrelevant operators; here
decoupling is not manifest
in contrast to the Lagrangian framework.

\setcounter{footnote}{0}
\bigskip
The canonical de Sitter space Green's function is built on the
preferred choice for the cosmological vacuum, the
Bunch-Davies/adiabatic state.
In the Hankel function $\cH_{\nu}$ basis of de Sitter mode-functions,
this vacuum is given by the choice $b(k)=0$:\footnote{
\label{fn:1}
Translating
back to Lagrangian effective field theory, eq. (\ref{eq:70}) shows that the BD-state
corresponds to a momentum-dependent effective boundary coupling $\kap_{BD}(k)$. The fact that the
the BD-state by definition reduces to the flat Minkowski vacuum for
high $k$ does mean that all irrelevant boundary operators are zero for
the BD-state. There are strong indications that the
BD initial conditions is an IR-fixed
point of boundary
RG-flow. (Qualitatively this is suggested by the fact that the
  BD state is the adiabatic vacuum. Quantitatively it is known that
  the BD-state is a fixed-point of the 'shift'-symmetry inherent in
  all boundary actions \cite{Schalm:2004qk}.)
This would mean that they are preferred initial conditions from an effective field
theory perspective as well. For an IR fixed point it is a
sensible procedure to study small departures by turning on irrelevant
operators (\ref{eq:50}). We proceed on this assumption.}
\begin{eqnarray}
\label{eq:130}
\vphi^{BD}_{+} (-k\eta) &=& (-k\eta)^{3/2} \sqrt{\frac{\pi}{4k}}\left(\frac{H}{k}\right)
{\bar{\cH}}_{\nu}(-k\eta)~,
\non
~~~
\vphi^{BD}_{-}(y) &=& (\vphi^{BD}_{+})^*(y)~,
~~~\nu = \sqrt{\frac{9}{4}-\frac{m^2}{H^2}}~.
\end{eqnarray}
In this basis departures from the Bunch-Davies state are directly
parameterized by a non-zero $b(k)$. Below we study small
irrelevant deformations of the BD initial conditions.

The non-zero contribution to $b(k)$ corresponding to the leading
irrelevant deviation $\delta \kappa = \beta {k^2 \over a_0^2 \,
M}$ from the BD initial state $\kap_0=\kap_{BD}$, can be obtained
by expanding
 (\ref{eq:70}) to lowest order
in $\del\kap$.
\begin{eqnarray}
\label{eq:100}
\delta b (k) = - \delta \kappa(k)  \left( {(\varphi^{BD}_{+,0})^2 \over
    \varphi^{BD}_{+,0} \left(\kap_{BD}\varphi^{BD}_{-,0} + \partial_n \varphi^{BD}_{-,0}\right)} \right) +
    \cO(\del\kap^2)~.
\end{eqnarray}
Using (\ref{eq:70}) (with $b_{BD}=0$) to solve for
$\kap_{BD}=-\pa_n\vphi^{BD}_+/\vphi^{BD}_+$,
the denominator in this expression can be rewritten using the Klein-Gordon normalization condition for the
dS mode functions $\varphi_{\pm}$
\begin{eqnarray}
\label{eq:110}
\varphi_{+}\, \partial_n \varphi_{-} - \varphi_{-}\, \partial_n \varphi_{+} = i a_0^{-3}~,
\end{eqnarray}
With these relations we obtain the following
expression for $\delta b(k)$
\begin{eqnarray}
\label{eq:120}
\delta b (k) = i a_0^3 \, (\varphi^{BD}_{+,0})^2 \, \delta \kappa = i a_0^3 \, (\varphi^{BD}_{+,0})^2 \,
\left( \beta {k^2 \over a_0^2 \, M} \right)~.
\end{eqnarray}
Note that the boundary coupling $\kap$ is independent of the choice of
basis $\vphi_{\pm}$ but $b$ and hence $\del b$ is not.
Let us also
emphasize that this expression can only be trusted for small $\delta
\kappa$, i.e. for boundary physical momentum scales $p_0 = k/a_0$
smaller than the cut-off scale $M$.

The change in the Green's function eq. (\ref{eq:13}) due to the
deviation
$\del b_{\del\kap}(k)$ from the BD state is now readily determined,
\begin{eqnarray}
  \label{eq:14}
\fl
  G_{\del b}(k,\eta_1,\eta_2) &=& G_0(k,\eta_1,\eta_2)
\non
&& + \Big[\Big(\del b \,
  \varphi^{BD}_{-}(\eta_1)\vphi^{BD}_-(\eta_2) + {\rm c.c.} \Big)\Big.
\non
&& ~~~\Big.+ 2 |\del b|^2 \,
  \varphi^{BD}_{-}(\eta_1)\varphi^{BD}_{+}(\eta_2)\Big]\theta_{12}
  +(\eta_1 \leftrightarrow \eta_2) + {\cal O}(\del
  b^3)
\non
&=& G_0(k,\eta_1,\eta_2)
\non
&& + \Big[\Big(\left(ia_0^3(\vphi^{BD}_{+,0})^2 \del
  \kap +
  \cO(\del\kap^2)\right)\varphi^{BD}_{-}(\eta_1)\vphi_-^{BD}(\eta_2)
+{\rm c.c.}\Big)
\Big.
\non
&& ~~~+ \Big. 2 a_0^6 |\vphi_{+,0}^{BD}|^4 |\del \kap|^2
  \vphi^{BD}_-(\eta_1)\vphi^{BD}_+(\eta_2)\Big]\theta_{12} + (\eta_1
  \leftrightarrow \eta_2) + \cO(\del\kap^3)~.
\end{eqnarray}
We expanded to {\em second} order in $\del b$. Contrary to
expectation --- as we will show in the next section --- it is the $a_0^6|\del\kap|^2$
term in the above equation whose contribution dominates the one-loop
gravitational backreaction. (We will not need to know the exact $\del
\kap^2 \vphi_-^2 +{\rm c.c.}$ terms, despite formally being of the same order.)

\section{Backreaction from initial state corrections}
\label{sec:backreacion}
\setcounter{equation}{0}

The one-loop backreaction we will calculate is formally a divergent
quantity. There are three
important points
to make in that
regard --- one general and two specific:
\begin{itemize}
\item[{\em a)}] In cosmological settings the Hadamard
  constraint that the stress-tensor
  can be rendered finite by subtracting the flat space counterterm
  has long been used as an initial state
  selection rule. In
 \cite{Schalm:2004qk} we argued that there are a large number of
 non-Hadamard initial conditions for which the stress tensor can
  be consistently renormalized.
  Here we will focus on the  specific
  irrelevant correction (\ref{eq:40}) to the Hadamard BD-state.

\item[{\em b)}] Wishing to discuss effects of new physics which
are encoded in irrelevant {\em non-renormalizable} operators, an
explicit cut-off is required to maintain finiteness of the theory.
`Renormalization' is the redefinition of all quantities under a
change of this cut-off, such that the new theory reproduces the
same physics (see e.g. \cite{Polchinski:1983gv}).

\item[{\em c)}] In
field-theoretic language the constraint that the expectation value of
the stress-energy tensor can be rendered finite
is equivalent to
renormalization of the composite operator $T_{\mu\nu}$.
Inherently any renormalization also needs a renormalization
prescription to
determine the finite remainder after the subtraction of
divergences. This prescription amounts to setting a predetermined
correlation function equal to an experimentally measured
quantity at a chosen scale.\footnote{Except for wavefunction renormalization, which is
  fixed by unitarity.} In a
theory with a boundary action the stress tensor consists of two parts;
a bulk and a boundary contribution.
Generically each will be divergent
and each will need a separate renormalization prescription. The
boundary stress tensor is naturally fixed by the value of $T_{\mu\nu}$
on the fixed conformal time boundary at $\eta_0$. This
quantity is beyond experimental reach, and the renormalization
prescription to use {will be unknown and ambiguous}. However, the
``boundary-energy'' encoded in the boundary stress tensor is
localized; the bulk physics far from the boundary is insensitive to
the particular boundary renormalization prescription. The theory is
predictive.
\end{itemize}
Recall also that 1) interactions are not relevant
for the one-loop contribution to the stress
tensor and that 2) we may
limit our attention to a massless scalar field as the high $k$ modes
running around the loop will dominate the answer (at the cost of
an IR divergence, which we should take care to isolate). With point
{\em c)} above in mind, the computation
therefore reduces to the expectation value of the bulk
stress tensor for
a free massless scalar
field w.r.t. the vacuum $|b\rangle$:
\begin{eqnarray}
\label{eq:160}
\langle b | T_{\mu\nu}(\eta) | b \rangle &=& \langle b |
T_{\mu\nu}^{bulk}|b\rangle ~,~~~~~ {\it if}~ \ln\left|\frac{\eta_0}{\eta}\right| \gg
  \frac{H}{M}~,
\non
T_{\mu\nu}^{bulk} &=&\partial_\mu \phi \partial_\nu \phi - \frac{1}{2}
g_{\mu\nu} \, g^{\rho\sigma} \partial_\rho \phi \partial_\sigma \phi~.
\end{eqnarray}
By neglecting the contribution of the boundary stress tensor, the
validity of this expression is intrinsically limited to
(physical) distances at least one
cut-off length $1/M$
away from the boundary (see e.g. \cite{Schalm:2004qk}). In conformal
time this implies the bound stated above.

Finally, in a homogeneous and isotropic background the two non-zero components
of the stress tensor are the density $T_{\eta\eta}= a^2(\eta) \rho$
and the pressure $T_{ij} = g_{ij}p $.
Expressed in terms of the useful quantities
\begin{eqnarray}
  \label{eq:17}
  K_{\mu\nu} \equiv \pa_{\mu}\phi \pa_{\nu} \phi~,~~~~\bar{K} =
  g^{ij}K_{ij}~,
\end{eqnarray}
they are
\begin{eqnarray}
  \label{eq:18}
\begin{array}{rcccl}
  a^{-2}T_{\eta\eta}  & =& \hlf\left(a^{-2}K_{\eta\eta}
  +\bar{K}\right) & =&   \rho~,
  \\[.1in]
  g^{\mu\nu}T_{\mu\nu} & =& a^{-2}K_{\eta\eta}-\bar{K} &=&
  -\rho+3p~.
\end{array}
\end{eqnarray}
 We are now ready to evaluate the one-loop
contribution to the stress-tensor. At one-loop the expectation
value of the quantities $a^{-2}K_{\eta\eta}$ and $\bar{K}$ is
given in terms of the {\em equal time} Green's function eq.
(\ref{eq:13})
\begin{eqnarray}
  \label{eq:19}
 \langle b| a^{-2}(\eta)K_{\eta\eta}(\eta)|b\rangle &=& \left. \int \frac{d^3
 k}{(2\pi)^3} \ove{a^2}\pa_{\eta_1}\pa_{\eta_2}
 G_b(k,\eta_1,\eta_2)\right|_{\eta_1=\eta_2=\eta} \non
\langle b| \bar{K}(\eta)|b\rangle &=& \int \frac{d^3 k}{(2\pi)^3}
 \frac{k^2}{a^2} G_b(k;\eta,\eta)~.
\end{eqnarray}
Expanding the Green's function around the BD-state to
second order as in eq. (\ref{eq:14}) each will have three
contributions.

\subsection{Zeroth order BD term}

The zeroth order term is the one-loop expectation value of
the stress-tensor w.r.t. the BD vacuum \cite{Birrell:ix}.
We will not
re-compute it here. What is important for us, is to note that the
naively divergent part is {\em independent} of $\eta$, if the regularization procedure
is so. This follows from the fact that the BD-state is Hadamard. This
constant divergence can be cancelled by the Minkowski space
counterterm plus a finite {\em constant} piece adjusted according to
the renormalization prescription. As is well known, there are
small finite time-dependent remainders. This is the `perturbative'
quantum instability of de Sitter space \cite{pertdS}.
Compared to the effect of the
irrelevant correction to the Bunch-Davies vacuum, these terms will
be negligible.

\subsection{First order term}

\begin{figure}[htbp]
  \centering
\includegraphics{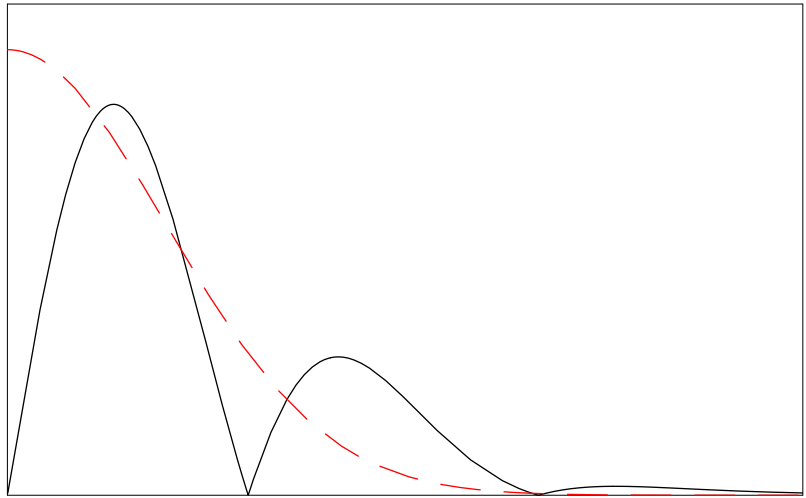}
  \caption{
Exact first order correction to stress-tensor due to the
irrelevant operator $\beta$ (absolute value of $\bar{K}$ in units of
the classical density $\rho=3H^2M_{pl}^2$; solid)
compared to the exponential scaling
$e^{-2M^2(t-t_0)^2}$ (dashed) derived in the high $k$ approximation.}
  \label{fig:1}
\end{figure}

A priori we expect the linear term in $\del
b$ to be the leading correction to
the stress tensor. The first order correction to the Green's function, however,
\begin{eqnarray}
  \label{eq:22}
\left.
  \Delta^{(1)}G(k;\eta_1,\eta_2)\right|_{\eta_1=\eta_2=\eta}  &=&
 \Big(ia_0^3\del\kap(\vphi^{BD}_{+,0})^2 (\varphi^{BD}_{-}(\eta))^2
+{\rm c.c.}\Big)
\end{eqnarray}
is a highly oscillating function in $k$. Indeed this
oscillatory behavior is a distinct characteristic of corrections to
the BD-vacuum, which may make them experimentally identifiable in the
inflationary power spectrum. The
backreaction, on the other hand,
corresponds to the integral of the Green's function over
$k$ and here the oscillatory peaks and valleys will largely cancel
each other out. For a massless field the integrals can be done
exactly, but the qualitative effect is clearly illustrated by
the dominant high $k$ part of the integral. In this limit the
mode functions simplify to
\begin{eqnarray}
  \label{eq:23}
  k \gg -\frac{1}{\eta} ~~~~~ \vphi_+(k,\eta) = -
  \frac{1}{\sqrt{2ka^2}} e^{ik\eta} +\cO(k^{-3/2})
\end{eqnarray}
resulting in a simple sine for the first order Green's function
\cite{Porrati:2004gz,Tanaka:2000jw}
\begin{eqnarray}
  \label{eq:16}
  k \gg -\frac{1}{\eta} ~~~~~ \Delta^{(1)}G(k;\eta,\eta) &=&
  ia_0\frac{\beta k^2}{M} \frac{1}{4k^2a_0^2a^2}e^{2ik(\eta_0-\eta)} +{\rm
  c.c} +\ldots \non
&=& \frac{\beta}{2Ma_0a^2}\sin(2k(\eta-\eta_0))+\ldots
\end{eqnarray}
In this high $k$ approximation the quantity $\bar{K}$, regulated with a
smooth cut-off function, evaluates to 
\begin{eqnarray}
  \label{eq:21}
  \bar{K} &\sim& \frac{\beta}{Ma_0a^4} \int dk k^4 \sin(2k(\eta-\eta_0)) e^{-k^2/2\cM^2}
\non
&  \sim&  \frac{\beta}{Ma_0a^4}\cM^5 e^{- 2
  \cM^2(\eta-\eta_0)^2}\Big(1 +\ldots \Big)~.
\end{eqnarray}
Note that the answer decreases in time only for a comoving cut-off
$\cM=a_0M$ rather than a physical one $\cM=a(\eta)M$; we will comment
on this below. The main point is that
in terms of physical time $t = -\ove{H}\ln |H\eta|$, $\bar{K}$
damps with scale $\cM/a_0 = M$.
\begin{eqnarray}
  \label{eq:24}
  \bar{K} &\sim & \beta M^4\frac{a_0^4}{a^4}
  e^{- 2\frac{M^2}{H^2}(1-\exp(-H(t-t_0))^2)} +\ldots
\non
&\sim & \beta M^4 \frac{a_0^4}{a^4}
  e^{- 2{M^2}(t-t_0)^2(1+ \cO(H(t-t_0)))} +\ldots
\end{eqnarray}
The exact answer for $\bar{K}$ (see
figure \ref{fig:1}) shows that all significant terms observe this
scaling behavior.\footnote{Because the Bessel function part of the
  mode functions
  simplify for a massless field ($H_{3/2} =-\sqrt{\frac{2}{\pi
      z}}e^{iz}(1+i/z)$) the exact answer with a Gaussian cut-off is readily obtained
  in terms of error-functions. Expanding around $M=\infty$, 
the
  first term which fails to scale as $e^{-M^2t^2}$ scales inversely
  with $M$:
$$
\bar{K}= \beta \left[e^{-M^2t^2}M^4\left(1+\cO(\frac{H}{M})\right) +
f(t)\frac{H}{M}\left(1+\cO(\frac{H}{M})\right)\right]+\cO(\beta^2)
$$
where $f(t)$ is a polynomial function in $Ht$.
} This result has the important consequence that the first order
backreaction, though non-vanishing, is essentially fully localized
on the boundary. Recall that in a low-energy effective theory with
cut-off scale $\cM$ all objects are smeared out over this scale:
all boundary objects effectively behave as a regulated 
distribution
 with scale $M$ --- Gaussian $Be^{-\cM^2(t-t_0)^2}$ or otherwise. We have therefore ventured outside the
range of validity of using only the bulk stress tensor to compute
the backreaction (\ref{eq:160}). The neglected formally infinite
boundary stress tensor and its counterterm contribute at first
order with the same scaling behavior. The first order term is
therefore almost completely fixed by the boundary renormalization
prescription, and has no significant bulk remnant. 

We conclude that
the most relevant term {\em for the bulk backreaction}  is the first
non-oscillatory contribution in the integral: the second order
$a_0^6\del\kap^2$ term mentioned below eq. (\ref{eq:14}). This had
already been surmised by Tanaka \cite{Tanaka:2000jw}.\footnote{Formally
this is of the same order as the first subleading irrelevant operator
correction $\beta_2k^3/M^2 (\phi_{+,0}^2\phi_-^2+{\rm c.c.})$. As is clear from eq. (\ref{eq:14}),
however, its contribution will also be oscillatory and
localized to the boundary.}
We note that in \cite{Porrati:2004gz} the first order term
is used to put constraints on the coefficient $\beta$. {Clearly
  first order constraints
are formally much stronger, but their time-dependence indicates
  that they are subject to a renormalization ambiguity.}

\subsubsection*{Which cut-off?}

Before we compute this leading non-oscillatory term, we give a
qualitative explanation why
the cut-off used ought to be $\cM=a_0M$ rather
than $\cM=a(\eta)M$. In general one is always free to
use whatever cut-off function one fancies, provided it renders all
Green's functions finite. In practice one chooses
one that respects the most
symmetries of the theory: this guarantees that the Ward identities are
manifestly obeyed in the regulated theory. A canonical example for a
regulator function is to modify the kinetic term to
\begin{eqnarray}
  \label{eq:26}
  - \hlf \int \pa_{\mu}\phi \pa^{\mu}\phi & \rar& -\hlf \int
    e^{-\frac{D_{\nu}D^{\nu}}{M^2}}  \pa_{\mu}\phi \pa^{\mu}\phi
\end{eqnarray}
where $D_{\mu}$ is the covariant derivative.
In a cosmological setting the metric inherent in the contraction
$D_{\mu}D^{\mu}$ introduces a time-dependence in the cut-off function
for the (spatial) momenta: $\exp(-\frac{D_{i}D^i}{M^2})=
\exp(k^2/a^2M)$. What matters for the computation above is how this
cut-off function affects the Green's function. Recall that the Green's
function is a bi-local function of two points, whereas the regulating
function is well-defined {(in terms of derivatives)} at one point. A qualitative way to
understand better what is happening is to not modify the kinetic term
explicitly, but rather implicitly through a change in the fields
\begin{eqnarray}
  \label{eq:25}
  \phi(x) & \rar & e^{-\frac{D_{\mu}D^{\mu}}{2M^2}}\phi.
\end{eqnarray}
This suggests  a (qualitative)
change for the cosmological mode functions
\begin{eqnarray}
  \label{eq:27}
  \vphi(k,\eta) &\rar & e^{-\frac{k^2}{2a^2M^2}} \vphi(k,\eta)
\end{eqnarray}
The
Green's function is therefore modified to
\begin{eqnarray}
  \label{eq:28}
  G(k;\eta_1,\eta_2) &\rar &
  e^{-\frac{k^2}{2a_1^2M^2}}G(k;\eta_1,\eta_2)
  e^{-\frac{k^2}{2a_2^2M^2}}
\end{eqnarray}
Inferring from the relation between the stress-tensor
expectation value and the
Green's function, $\langle b| T_{\mu\nu}(\eta)|b\rangle \sim
\pa_{\mu_1}\pa_{\nu_2} G(k;\eta_1,\eta_2)|_{\eta_1=\eta_2}$, one is
inclined to use a naive regulating function $e^{-k^2/a^2M^2}$ in
eq. (\ref{eq:21}). This
is {\em not} correct. When the change
in the boundary condition is treated as a perturbation, the first
order term involves not one but two Green's functions; each interpolates between a
boundary vertex and the location of the composite operator. This is
clearly shown in \cite{Schalm:2004qk} and is indicated by the presence
of four mode functions in eq. (\ref{eq:22}). The counterintuitive
 answer obtained for $\bar{K}$ with a physical
 cut-off $\cM=a^2M^2$ is a manifestation of using this wrong cut-off.

According to the above, it is clear that
the regulating function to be used in
eq. (\ref{eq:21}) is
\begin{eqnarray}
  \label{eq:29}
  \cF_{reg} = e^{-\frac{k^2}{a^2M^2}-\frac{k^2}{a_0^2M^2}} =
  e^{-\frac{k^2}{a_0^2M^2}(\frac{a^2/a_0^2+1}{a^2/a_0^2})} \simeq e^{-\frac{k^2}{a_0^2M^2}}~.
\end{eqnarray}
The time-dependence in the ratio $\frac{a^2/a_0^2+1}{a^2/a_0^2}$ which
 smoothly interpolates between two and unity is clearly not relevant.

\subsection{Second order term}

Let us now proceed and calculate the dominant, non-oscillatory, second order contribution
to the backreaction (\ref{eq:160}). In the high $k$ approximation
(\ref{eq:23}) for the mode functions
the integral can easily be performed and after some straightforward
algebra, we find
\begin{eqnarray}
\label{eq:190}
\bar{K}^{(2)} &=& {1 \over 3 (4\pi)^2} \, |\beta|^2
\, M^4 \, e^{-4H(t-t_0)}
\left[1+\cO\left(e^{2H(t-t_0)}\frac{H^2}{M^2};\frac{H}{M}\right)\right]~\non
a^{-2}{K}^{(2)}_{\eta\eta} ~&=& {1 \over 3 (4\pi)^2} \, |\beta|^2
\, M^4 \, e^{-4H(t-t_0)} \left[1+\cO\left(e^{2H(t-t_0)}\frac{H^2}{M^2};\frac{H}{M}\right)\right]~.
\end{eqnarray}
The equality of the two leading terms, implying $\rho=3p$,
reflects the absence of a scale
in a free massless field. They differ of course
at the subleading level, as
an exact calculation shows.

The time behavior of the answer is fundamentally different from the
first order term (\ref{eq:24}). This can be
attributed to the non-oscillatory behavior of the integrand.
The backreaction still decays exponentially fast away from the position
of the boundary, but now with the Hubble scale $H$ rather than the
cut-off scale $M$. As
this contribution is clearly unaffected by both the
renormalization prescription of the bulk stress tensor (a constant
time-independent term) and the boundary stress tensor (a localized
term scaling with the cut-off $\cM$), it unmistakably corresponds to a
physical contribution. Hence an irrelevant perturbation away from the
BD-state introduces an excess amount of physical energy density \cite{cos-vac-SK}.
As time progresses this energy density red-shifts due to the dS
exponential expansion, explaining the exponential decay with scale
$H$. Because this contribution entails a physical change in the
background, the observed cosmology
constrains the size the coefficient $\beta$ of the
irrelevant operator. The
question is whether these constraints (dis)allow potentially
observable corrections to the CMB.

\subsection{Adiabaticity and initial conditions}
\label{sec:adiab-init-cond}

Aside from determining phenomenological constraints on possible
high-energy physics effects in the CMB, the calculation of the
one-loop backreaction elucidates a theoretical argument as well.
By consensus the preferred vacuum state in
cosmological settings is the adiabatic one. This state is defined
as that state in which the number operator decreases slowest with
time. This is a somewhat uncomfortable definition as the number
operator is not an observable quantity. Its observable
counterpart, however, is the normal-ordered Hamiltonian, i.e. the
renormalized stress tensor. An improved definition of the
adiabatic initial conditions is those for which the (one-loop)
stress-tensor changes slowest in time. As our calculation clearly
shows, any irrelevant perturbation away from the BD-initial
conditions will introduce a scaling $e^{-nH(t-t_0)}$. Hence we
recover that the BD-initial conditions are the adiabatic ones.
Moreover irrelevant corrections to the BD-state quantitatively
parameterize deviations away from adiabaticity.

This quantitative understanding of non-adiabatic initial states
(indeed the presence of irrelevant operators means that the theory is
not renormalizable and decoupling no longer works exactly)
 and
the consequent time-dependence of the stress-tensor illustrates an
interesting new contribution to the standard slow-roll
inflationary scenario (see the recent article
\cite{Danielsson:2004xw} for a qualitatively similar viewpoint). As a variant of
one-loop induced inflationary potentials, the above computation shows
how non-adiabaticity can be responsible for deviations from the
classically expected scale-invariant de Sitter spectrum. Such
non-adiabatic contributions to inflationary evolution
are arguably natural in the
interpretation of cosmological evolution as relaxation towards a
ground state. Experimental evidence, however, ought to judge the
full validity of such a `relaxation'-scenario. In the next section we
will discuss to what extent deviations from adiabaticity, i.e. the
contribution to the power spectrum from boundary irrelevant operators,
contribute to the inflationary evolution and are constrained by
current observation.

\section{Constraints from backreaction}
\label{sec:constraints}
\setcounter{equation}{0}

These constraints will follow from comparing the
measured vs. predicted gravitational background. We have already shown
the results of this
calculation in the right panel of Figure \ref{fig:2} where we compare
them with the direct observational bounds from the power spectrum plus
a constraint on the maximal value of $H/M$. Let us now show, how we
arrived at the bounds in Figure \ref{fig:2}.

The Friedmann equation relates the measured full quantum corrected
Hubble scale $H_{eff}$ to the expectation value of the stress-tensor
\begin{eqnarray}
\label{eq:200}
\fl
H_{eff}^2(\eta) = {1\over 3M_p^2} \langle T_{00} \rangle =
\ove{3M_p^2} \left( T_{00}^{(0;0)}(\eta) + T^{(1;0)}_{00}(\eta)+
  T^{(1;1)}_{00}(\eta)+ T^{(1;2)}_{00}(\eta)+\ldots\right)~,
\end{eqnarray}
Here $T_{00}^{(n;m)}$ corresponds to the $n$-loop order $\beta^m/M^m$
irrelevant
corrections to the
backreaction, and $M_p$ equals the reduced Planck mass $M_p \approx
2.4 \cdot 10^{18}$ GeV. We have shown earlier that $T^{(1,1)}_{00}$ is
localized on the boundary and, concentrating only on the leading $M^4$
contributions, that the $T^{(1,0)}_{00}$ term is constant at this
  order and fixed by the renormalization prescription. We
  will therefore ignore their contributions from here on. $T^{(0,0)}$
  is of course the classical background and at leading
  order $M^4$, $T^{(1,2)}_{00}$ follows from (\ref{eq:190})
  \begin{eqnarray}
    \label{eq:33}
    T^{(1,2)}_{00} = \frac{1}{48\pi^2}|\beta|^2M^4 e^{-4H_0(t-t_0)}
  \end{eqnarray}
with $H_0$ the classical Hubble scale.

The first constraint is simply that of a consistent perturbation
theory, i.e. that the one-loop backreaction term $T^{(1,2)}_{00}$
should be
small compared to the measured expansion rate of the universe
\begin{eqnarray}
\label{eq:210}
{T^{(1,2)}_{00} \over T^{eff}_{00}} \ll 1~~~\Rightarrow~~~|\beta|^2 e^{-4H_0(t-t_0)}\ll (12 \pi)^2 \left({M_p^2 H_{eff}^2
\over M^4}\right)~.
\end{eqnarray}
 This constraint is strongest at the `earliest time' $t \sim t_0$ where we
 set the initial conditions. We see that $|\beta|^2 \ll (12\pi)^2
 M_p^2H^2/M^4$.
In conventional string scenario, the string scale is roughly two
orders of magnitude below the (reduced) Planck scale which gives
$|\beta|^2 \ll 10^3$ for $H \sim 10^{14}$ GeV.
 This establishes our claim that the backreaction
constraint has no practical content for $H/M >10^{-4}$. With the
numbers used for $H$ and $M$ the power spectrum already constrains
$\beta \leq 1$. The reason why this is so, is also evident. The
weakness of the backreaction constraint is a result of the
cancellation of the highly oscillatory integrand. The
power spectrum, however, is roughly the integrand of the
one-loop backreaction. For the power spectrum, the oscillations do
not cancel each other, which explains the higher sensitivity
compared to the backreaction constraint.

In addition to this `static' backreaction constraint, one should also
demand that the
time derivatives of the backreaction are not too big. These are the
phenomenological parameters that determine the characteristics of the
inflationary evolution. Through one-loop backreaction the boundary
irrelevant operators will contribute to these, as we anticipated in
section \ref{sec:adiab-init-cond}. The derivatives must remain small
enough, however, to
guarantee that inflation lasted long enough and to
explain the measured
scale-invariance of the spectrum.  The (first)
time-derivative of the stress tensor is related to the conventional
inflationary slow-roll parameter
$\epsilon \equiv - \frac{\dot H}{H^2}$ as $ H\dot{T}_{00}/T_{00}=
- 2\eps$. To guarantee inflation
$\eps$ should be smaller than $1$
\begin{eqnarray}
\label{eq:220}
-\frac{\dot{T}_{00}}{6H_{eff}^3M_p^2} \equiv \eps_{eff}
\ll 1 ~~~\rar ~~~
\frac{1}{2(6\pi)^2}|\beta|^2\frac{M^4}{M_p^2H^2}e^{-4N}
\simeq \eps_{eff}-\eps^{(0)}~.
\end{eqnarray}
For the classical de-Sitter background used here $\eps^{(0)}=0$
which implies the following constraint on $\beta$ (again evaluated close to the boundary)
\begin{eqnarray}
\label{eq:230}
|\beta|^2 \lesssim 2 (6\pi)^2 \left( {M_p^2 H_0^2 \over M^4} \right) |\epsilon_{eff}|~.
\end{eqnarray}
Using the same estimates as before, plus the experimental value for $\epsilon_{eff} \sim 10^{-1}$ we
conclude that $|\beta|^2 \lesssim 10^2$, which is again superseded by the
power spectrum sensitivity itself.

Finally the measured scale invariance of the power spectrum constrains
the second time derivative of the backreaction.\footnote{Because we only have observational knowledge
  about the first and second order slow roll coefficients, higher derivatives
  $d^nT_{00}/dt^n$ place no phenomenological constraints on the value
  of $\beta$. An observed running of the spectral index would bound
  the third derivative as well.} It is related
to a combination of the first and the second
slow-roll parameter $\frac{\ddot{T}_{00}}{H^2T_{00}}= 2\epsilon \eta
+2\eps^2$. We deduce
\begin{eqnarray}
\label{eq:240}
\fl
\hspace{1.7in}
\frac{\ddot{T}_{00}}{2H^2T_{00}} \equiv
\eps_{eff}(\eta_{eff}+\eps_{eff})&& ~~~ \non
\fl
\rar~~~
{1 \over 2(3\pi)^2} |\beta|^2  e^{-4N}\, \left(
  {M^4 \over M_p^2 H_0^2}\right) \simeq
\eps_{eff}(\eps_{eff}+\eta_{eff})&&- \eps^{(0)}(\eta^{(0)}+\eps^{(0)})
\end{eqnarray}
For a general slow-roll inflationary background this will lead to the following
constraint on $\beta$,
\begin{eqnarray}
  \label{eq:31}
   |\beta|^2  \, \lesssim \, (6\pi)^2\left(\frac{M_p^2 H_0^2}{M^4}\right)  \eps_{eff}(\eps_{eff}+\eta_{eff})~.
\end{eqnarray}
This is clearly our strongest constraint. Using the same estimates as
  before and in addition
assuming that $|\epsilon_{eff}| \sim 0.1~,~|\eta_{eff}| \sim 0.1$, we
  approximately
find $|\beta|^2 \ll 10$, but this number could vary somewhat depending
  on the precise estimates.
Yet the power spectrum constraint is again stronger.

Eqs (\ref{eq:210})-~(\ref{eq:31}) are the bounds we have drawn in the
right panel of figure \ref{fig:2}. For comparison we have also shown
the order of magnitude estimates made in
\cite{Porrati:2004gz}. Mathematically the latter do correspond to the
first order one-loop correction to the backreaction. We have shown,
however, how this contribution is localized on the boundary where one
sets the initial conditions. It is therefore subject to a
renormalization ambiguity. The second order contribution is physical
but its constraints are a lot
milder and are not
threatening the potential observability of initial state corrections
in the CMB. They clearly leave a large window of opportunity for a
significant range of values of $H/M$ where irrelevant corrections will
affect the CMB.

\subsection{The length of inflation and initial conditions}
Precisely the fact that non-Bunch-Davies initial
  conditions ought to correspond to a quantum physical contribution to
  the stress tensor
  has been used as a qualitative argument that irrelevant corrections
  to the initial conditions cannot be large enough to be potentially
  observable. The energy present in the initial state would {\em
  blueshift} towards the past and rapidly invalidate the classical
  inflationary background used. Since we need roughly sixty e-folds of
  inflation to explain the observed flatness and isotropy  of our
  universe this would be problematic. With the quantitative boundary
  effective formalism to encode the initial conditions we can
  investigate this issue more deeply. Indeed the
scaling with $H$ of the second order correction to the stress tensor
  implies
that for nonzero $\beta$ there is an excess
physical energy present on top of the classical background. This
  backreaction starts to dominate over the classical background when
\begin{eqnarray}
  \label{eq:34}
  e^{4N} = (12\pi)^2 \left(\frac{M_p^2H^2}{M^4|\beta^2|}\right)
\end{eqnarray}
with $N=H(t_0-t)$ the total number of e-folds {\em before} the
`earliest time'. Only for very small values of $H/M$ is this number
\begin{eqnarray}
  \label{eq:1}
  N  \simeq \frac{21}{4} - \hlf\ln|\frac{\beta M}{H} |~.
\end{eqnarray}
negative. For the range of values $H/M$ of interest the time
where backreaction starts to dominate is therefore {\em before} the `earliest
time' where one can start to trust the cosmological low energy
effective action. What truly happens before the `earliest time' is
unanswerable within the low energy effective framework. This does not
mean that the energy that is stored in the initial state miraculously
disappears or is not an backreaction issue before time $t=t_0$. It
does mean that we need to know the short distance completion of the
theory to answer this criticism. Within the range of validity of the
effective action after the `earliest time' the backreaction is always
a small perturbation on the classical background. The cosmological low
energy effective action with boundary is consistent.

A separate criticism argued that irrelevant corrections to the
initial conditions were necessarily fine tuned. We will answer
this elsewhere; a first estimate is made in \cite{porrati2}.
One noted fine
tuning objection: that any
momentum dependent feature in the power spectrum must be fine
tuned as we only measure a narrow ten $e$-fold window of the full
power spectrum produced during inflation, does not hold here.
It is a straightforward
exercise to show that any irrelevant operator
will introduce a power dependence in the
spectrum. Moreover, following the precepts of effective action this
momentum dependence occurs in a well-defined derivative expansion
which predicts outwardly differing but intrinsically universal
results for any ten e-fold window of the power spectrum.

\section{Conclusions}
\label{sec:conclusion}
\setcounter{equation}{0}

The main conclusion stemming from this work is that the
unambiguous backreaction
constraints on the
irrelevant operator coefficient are mild and do not affect the
potential observability of
the initial state corrections in the CMB. The three constraints
arising from the observed slow-roll inflationary phase read
\begin{eqnarray}
\label{eq:260}
(1) \quad |\beta|^2 &\ll& (12 \pi)^2 \, \left({M_p^2 H_0^2 \over M^4}\right)~, \\
(2) \quad |\beta|^2 & \lesssim & 2 \, (6 \pi)^2 \, |\epsilon_{eff}| \,
(\left( {M_p^2 H_0^2 \over M^4} \right)~, \\
(3) \quad |\beta|^2 & \lesssim & (6 \pi)^2 \, |\epsilon_{eff}| \,
(|\eta_{eff}| \, \left( {M_p^2 H_0^2 \over M^4} \right)~.
\end{eqnarray}
Typical, although slightly optimistic, estimates imply that ${M_p H_0}/{M^2} \gtrsim 1$ and $\epsilon_0
\sim \eta_0 \lesssim 10^{-1}$. The constraints on $|\beta|^2$ thus range from ${\cal O}(10^3)$,
${\cal O}(10^2)$ to ${\cal O}(10)$. These numbers could easily change
by an order of magnitude by slightly changing the precise parameter estimates, so there is some
room
for adjustment
in these constraints. This is plotted in Figure \ref{fig:2}.
Nevertheless, using the
suggested parameter estimates, order one coefficients for $\beta$ are still allowed by these
constraints. Indeed
the observability of the initial state corrections in the
CMB is not ruled out. Rather the power spectrum itself places stronger
constraints on $\beta$.

We have moreover shown that the
irrelevant corrections to the boundary effective action encoding
the initial conditions parameterize deviations from adiabaticity.
They correspond to having an excited energetic initial state.
Though this energy could cause a problem when blueshifted to the
past, within the framework of the low energy effective action it
is always perturbative. This ensures consistency of the
cosmological low energy effective action including high energy
induced irrelevant corrections to the initial state.

The burning question is whether these short distance corrections
are actually decipherable from CMB data. Similar to earlier
predictions of the contribution to the power spectrum, we find
that the leading irrelevant boundary operator yields a
characteristic oscillatory signature. In principle this ought to
make these corrections uniquely identifiable in the data. Sensitivity
studies in a specific UV-complete framework appear to bear this out \cite{Easther:2004vq}. As
figure \ref{fig:2} shows, however, the leading irrelevant correction
alone induces a very rapid oscillation period for those modes where
the amplitude is large enough. 
From eq. (\ref{eq:3c}) we see that its frequency is about
$\ome \simeq \frac{y_0}{2\pi} \simeq \frac{kM}{2\pi k_{max}H} \sim
\frac{M}{H}$. This is probably too rapid to be resolved in the
actual data.
It is difficult
to make a concrete statement, though. In the end only a full
deconvolving of the actual measured data with respect to the
template provided here, can yield an answer to the question
whether the accuracy with which the data is collected can uniquely
identify the contributions from short distance corrections to the
initial state. This is a calculation we intend to return to.

Let us finally mention that it would be extremely interesting if
one could calculate the coefficients $\beta$ from first
principles. For this we need to know the UV completion of the
theory.  For instance, it could be interesting to study some of the
recently proposed string inflationary scenarios \cite{KKLT} to see
whether one could explicitly calculate the coefficients of some of
the irrelevant operators in this context. With a UV-completion in
hand, one could moreover compute the unambiguous answer for the
first-order backreaction. If that were possible
one could start making concrete predictions for stringy signatures
in the CMB spectrum, a particularly exciting prospect.

\ack
It is a pleasure to thank Jan de Boer, Daniel Chung, Csaba Cs\'{a}ki,
Richard Holman,
Emil Mottola,
Bob McNees,
Massimo Porrati and Larus Thorlacius for interesting discussions and comments. GS and JPvdS are
particularly grateful to the organizers and participants of the MCTP
Dark side of the Universe May 2004
workshop and those of the Strings at CERN July 2004 workshop. KS and
GS would like to thank the organizers
and participants of the Banff New Horizons in String Cosmology June
2004 workshop. BRG and KS acknowledge financial support from
DOE grant DE-FG-02-92ER40699. The work of GS was supported in part by
NSF CAREER Award No. PHY-0348093, DOE grant DE-FG-02-95ER40896 and a
Research Innovation Award from Research Corporation.

\end{document}